\documentclass[twocolumn,showpacs,preprintnumbers,nofootinbib,prd,superscriptaddress,groupedaddress,10pt]{revtex4-1}

\usepackage{graphicx,amssymb,amsmath,amsthm,amsfonts,epsfig,epsf}
\usepackage[linktocpage]{hyperref}
\usepackage[usenames]{color}
\usepackage{epstopdf}

\usepackage{aas_macros}
\usepackage{bm}
\usepackage{dcolumn}
\usepackage[latin1]{inputenc}
\usepackage{latexsym}
\usepackage{rotating}
\usepackage{hyperref}
\usepackage{color}
\usepackage{longtable}
\usepackage{enumerate}
\usepackage{tensor}
\usepackage{url}
\newcommand{\ben}{\begin{enumerate}}
\newcommand{\een}{\end{enumerate}}

\def\be{\begin{equation}}
\def\ee{\end{equation}}
\def\bea{\begin{eqnarray}}
\def\eea{\end{eqnarray}}

\newcommand{\beq}{\begin{eqnarray}}
\newcommand{\eeq}{\end{eqnarray}} 
\newcommand{\ba}{\begin{align}}
\newcommand{\ea}{\end{align}}

\begin{document}

\title{Collapsing shells, critical phenomena and black hole formation}

\author{
Vitor Cardoso$^{1,2,3}$, 
Jorge V. Rocha$^{4}$
}
\affiliation{${^1}$ CENTRA, Departamento de F\'{\i}sica, Instituto Superior T\'ecnico -- IST, Universidade de Lisboa -- UL,
Avenida Rovisco Pais 1, 1049 Lisboa, Portugal}
\affiliation{${^2}$ Perimeter Institute for Theoretical Physics, 31 Caroline Street North
Waterloo, Ontario N2L 2Y5, Canada}
\affiliation{${^3}$ Dipartimento di Fisica, ``Sapienza'' Universit\`a di Roma \& Sezione INFN Roma1, P.A. Moro 5, 00185, Roma, Italy}
\affiliation{${^4}$ Departament de F\'isica Fonamental, Institut de Ci\`encies del Cosmos, Universitat de Barcelona, Mart\'i i Franqu\`es 1, E-08028 Barcelona, Spain.}

\begin{abstract}
We study the gravitational collapse of two thin shells of matter, in asymptotically flat spacetime or constrained to move within a spherical box. We show that this simple two-body system has surprisingly rich dynamics, which includes prompt collapse to a black hole, perpetually oscillating solutions or black hole formation at arbitrarily large times.
Collapse is induced by shell crossing and the black hole mass depends sensitively on the number of shell crossings.
At certain critical points, the black hole mass exhibits critical behavior, determined
by the change in parity (even or odd) of the number of crossings, with or without mass-gap during the transition. 
Some of the features we observe are reminiscent of confined scalars undergoing ``turbulent'' dynamics.
\end{abstract}



\maketitle

\noindent{\bf{\em I. Introduction.}}
The advent of supercomputers and the ability to solve numerically the Einstein field equations
has widened our knowledge of gravitational physics, and in some cases opened completely new and unexpected
directions~\cite{Cardoso:2014uka,Dias:2015nua}. Black hole (BH)
physics remains, in this context, one of the most challenging problems due to 
the large amplitude of the gravitational potential close to
their ``surface'' and because most often than not,
BHs interact at high energies. In recent years, important strides have been taken.
The two-body problem was solved satisfactorily at a numerical level~\cite{Cardoso:2014uka,Lehner:2014asa}, allowing for 
an understanding of binary BHs, including the violent collision
of two BHs at close to the speed of light~\cite{Sperhake:2008ga,Sperhake:2009jz,Sperhake:2012me,Healy:2015mla}.

In parallel, the question of whether and how BHs form as a result of the time-evolution of initial data was addressed in a seminal work by Choptuik~\cite{Choptuik:1992jv}. Choptuik studied the collapse of massless fields in asymptotically flat spacetime, finding critical behavior at the onset of BH formation, and dispersal of the scalar for very low amplitudes~\cite{Gundlach:2002sx}, consistently with the nonlinear stability of Minkowski~\cite{1993gnsm.book.....C}.

Only recently, however, did one start to understand the onset of the rich dynamics allowed for when the fields
are constrained to a finite spatial extent~\cite{Bizon:2011gg,Buchel:2012uh,Balasubramanian:2014cja,Craps:2015jma,Abajo-Arrastia:2014fma,Okawa:2014nea,Okawa:2015xma,Olivan:2015fmy}. Some classes of initial data seem to {\it always} collapse to BHs irrespective of their amplitude, although BH formation may take arbitrarily large times to occur; other types of initial data form nonlinearly stable, regular but oscillating configurations. The exact nature and development of the process is not known, but blue- and red-shifting of the radiation encoded in the nonlinearities of the field equations, are of paramount importance. Equally important is the confining nature of the setup, forcing the fields to slosh back and forth, allowing nonlinearities to build up.

The study of gravitational collapse of matter involves state-of-the-art numerical codes able to follow the system accurately for 
a large amount of time. The purpose of this {\it Letter} is to point out that similar phenomena occurs in a very simple scenario, indeed perhaps the simplest two-body problem one can conceive of: two infinitely-thin, spherically symmetric shells of matter in a box.
Such configuration is not only of astrophysical interest, but has been used many times in the past to understand fundamental problems in gravity~\cite{Israel:1966rt,Dray:1985yt,Nunez:1993kb}.

 This system possesses some of the features one would look for when trying to understand gravitational collapse: the two shells interact gravitationally and can exchange energy (leading to blue- or redshift effects in the shells); they are forced to either interact or collapse, a feature present also in anti-de Sitter (AdS) spacetime; and most importantly, the dynamics is easy to solve for.

Depending on the initial conditions, we find perpetually oscillating solutions, prompt collapse to BHs and even BH formation at arbitrarily large times. 
We are not implying that, say, the full complexity of gravitational ``turbulent'' phenomena is present in our setup; we are merely showing that aspects of the problem of gravitational collapse in confining geometries are present in this simple setting, which can be formulated simply and whose resolution involves only {\it first order ODEs}. This makes it easy to generalize to other frameworks (higher dimensions, theories with a cosmological constant, rotation, etc.) and to isolate the important features. In addition, we also show that even ``free'' shells moving in an asymptotically flat spacetime can exchange energy in such a way that the system never collapses nor disperses, forming instead a bound, oscillating configuration.

\noindent{\bf{\em II. Setup.}}
We focus exclusively on spherically symmetric spacetimes. We consider pure General Relativity
with no cosmological constant, but our results and methods are easily generalizable.
On this spacetime, we set up a confining box of radius $R_{\rm ext}$ which reflects all incoming matter.
Inside the box, we add two infinitely-thin, spherically symmetric shells of matter each of radius $R_{1,2}$.
The motion of the shells is governed by Einstein equations, which in this case can be solved at full nonlinear level, since inside and outside the shell the solution has the Schwarzschild form~\cite{Israel:1966rt} (see~\cite{Lake:1979zz} for the case of more general spherically symmetric geometries).

Consider first a single shell which will, generically, be described by the Schwarzschild solution
with mass $M_-$ and whose exterior is also Schwarzschild with mass $M_+$. The shell's radius is $R(\tau)$, where $\tau$ denotes the shell's proper time. The induced metric on the shell is 
\be
d\sigma^2=-d\tau^2+R(\tau)^2d\Omega^2\,,
\ee
and the non-vanishing components of the extrinsic curvature are (with $\pm$ exterior and interior, respectively)
\beq
K_{\tau\tau}^\pm&=&-\frac{\dot{\beta}_\pm}{\dot{R}}\,,K_{\theta\theta}^\pm=R\beta_\pm\,,K_{\phi\phi}^\pm=R\sin^2\theta\beta_\pm\,,\\
\beta_\pm     &\equiv&\sqrt{\dot{R}^2+1-2M_\pm/R}\,.
\eeq
Here, the overdot denotes a derivative with respect to $\tau$.

The shell is taken to be described by the following surface stress-energy tensor,
\be
S_{ij}=(\rho+P)u_iu_j+Pg_{ij}\,,
\ee
where $u_i$ is the perfect fluid's $3-$velocity, $\rho$ its energy density and $P$ its pressure.
Defining the jump of a given quantity $X$ across the shell surface as
$[X]\equiv X_{+}-X_{-}$, the Israel-Darmois conditions~\cite{Darmois:1927,Israel:1966rt}
\be
8\pi G S_{ij}=-\left([K_{ij}]-g_{ij}[K]\right)\,,
\ee
where $K_{\pm}=\dot{\beta}_+/\dot{R}+2\beta_\pm/R$ is the trace of the extrinsic curvature, yield
\beq
\beta_+-\beta_-+4\pi G R\rho=0\,,\label{eqbeta}\\
P+\frac{\rho}{2}+\frac{d}{dR}\left(\frac{R\rho}{2}\right)=0\,.\label{eqrho}
\eeq
Assuming, for simplicity, that the shell is described by an equation of state
of the form $P=w\rho$, with $w$ a constant, we find from \eqref{eqrho} that
\be
\rho=\frac{m}{4\pi G R^{2+2w}}\,,
\ee
with $m$ a constant, corresponding to the shell's invariant mass and $G$ denoting Newton's constant. 
Inserting this solution in \eqref{eqbeta}, we finally have
\be
\dot{R}^2+V=0\,,\label{eqmotion}
\ee
where the effective potential is
\be
V=1-\frac{M_++M_-}{R}-\frac{(M_+-M_-)^2}{m^2}R^{4w}-\frac{m^2}{4R^{2+4w}}\,.\label{potential}
\ee
Notice that the shell material obeys all the relevant energy conditions~\cite{Hawking:1973uf}, in particular
the null ($\rho+P\geq 0$), the weak ($\rho\geq 0, \rho+P\geq 0$),
the strong ($\rho+P\geq 0, \rho+3P\geq 0$) and the 
dominant ($\rho\geq P\geq-\rho$) energy conditions, for any positive density
$\rho$ and $1\geq w\geq-1/3$.

\begin{figure}[h!]
\begin{center}
\includegraphics[width=0.5\textwidth]{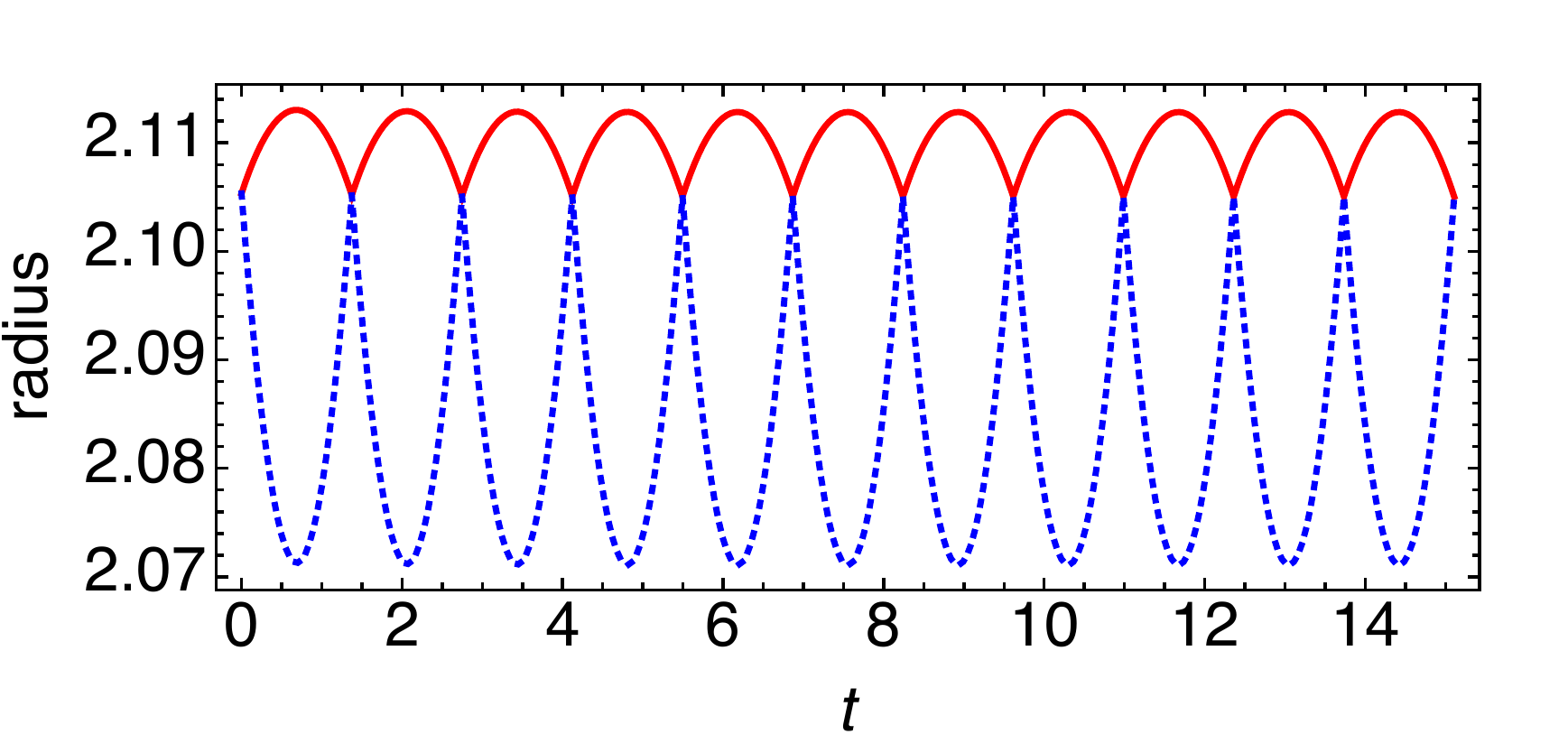}
\end{center}
\caption{Evolution of a ``free'' double-shell system under their own gravitational field.
The solid (red) curve denotes the outermost shell and dotted (blue) points the innermost.
We take the total gravitational mass $M_1=1$ , and the equation of state to be specified by
$w=0.86$. Both shells have identical mass parameter $m=2.9$; the inner gravitational mass was tuned to
$M_2=0.663$. The crossing point is at $R=2.10533719$.}
\label{fig:double_sustained}
\end{figure}

Equation \eqref{eqmotion} allows us to follow the position of the shell, given the amount of matter on the inside.
Algorithmically thus, the two-shell problem is solved by using \eqref{eqmotion} to track the position $R_{1,2}$
of the outer- and innermost shells: the innermost shell is described by \eqref{eqmotion} with $M_-=0,M_+=M_2$ and the outermost by
$M_-=M_2,M_+=M_1$. When the outermost shell reaches the box at $r=R_{\rm ext}$, it is perfectly reflected ($\dot{R}_{1}\to-\dot{R}_{1}$). 

We must also decide what happens whenever the two shells collide. Thin shells of dust would simply cross without any interaction---besides the gravitational influence, which is already taken into account by the junction conditions. However, we are interested in endowing our shells with non vanishing pressure, thus counteracting each shells' gravitational self-attraction. Depending on initial conditions, on each journey towards the center a shell can then either collapse or bounce back. Nevertheless, we model the problem by assuming that the two shells cross without 
changing their invariant masses and taking their 4-velocities to be continuous at the collision. This, however, does {\em not} preclude the shells from exchanging energy~\cite{1999PThPh_v2}. With this physically-sensible assumption, after the shells cross the gravitational mass $M_2'$ exterior to the first shell can be read off from Eq.(3.18) in Ref.~\cite{1999PThPh.101..989I} (see also Refs.~\cite{Nunez:1993kb,1999PThPh_v2}), which is obtained from conservation of energy and momentum during the collision, assuming the shells are ``transparent''.

Because of the immense freedom in the problem (in choosing the initial location of each shell, the equation of state parameter $w$, the mass parameter $m$ and masses of each shell), we will focus exclusively on shells initially on the same location $R(t=0)=R_i$, with the outermost shell expanding and the innermost contracting, and on the following two sets of initial conditions~\footnote{These initial conditions {\it were not} fine-tuned, and parameters close to these yield the same qualitative behavior. Thus, the behavior we discuss in the following is characteristic of a generic class of configurations. However, there are certain ranges of parameters, for example very small $w$, for which we observe prompt collapse only.}:

\noindent {\bf A.} $M_1=1, M_2=0.5,\,m_1=m_2=0.9,\,w_1=w_2=0.2$. These are shells of equal invariant mass and equal equation of state, and the free parameter in this set is the initial location of the shells, $R_i$. 

\noindent {\bf B.} $M_1=\delta, M_2=0.1\delta,\,m_1=m_2=0.1\delta,\,w_1=w_2=1,\,R_i=1$. The free parameter here is $\delta$ and quantifies the energy content in the spacetime.

The numbers above refer only to the initial conditions, as the mass of the innermost shell, $M_2$, varies during each crossing.
The formation of an horizon is signaled by the appearance of a zero of $1-2M_{1,\,2}/R_{1,\,,2}$. Typically (but not always), an horizon forms first at the location of the inner shell.

\noindent{\bf {\em III. Free oscillating shells.}}
This simple setup displays nontrivial dynamics, triggered by the energy exchange during each crossing of shells
and the mutual gravitational interaction.
For example, we are able to find solutions in asymptotically flat spacetime (i.e, without the confining box) describing two oscillating shells with identical invariant mass and equation of state. Such oscillating solutions do not exist for a single shell. 

\begin{figure}[h!]
\begin{center}
\includegraphics[width=0.5\textwidth]{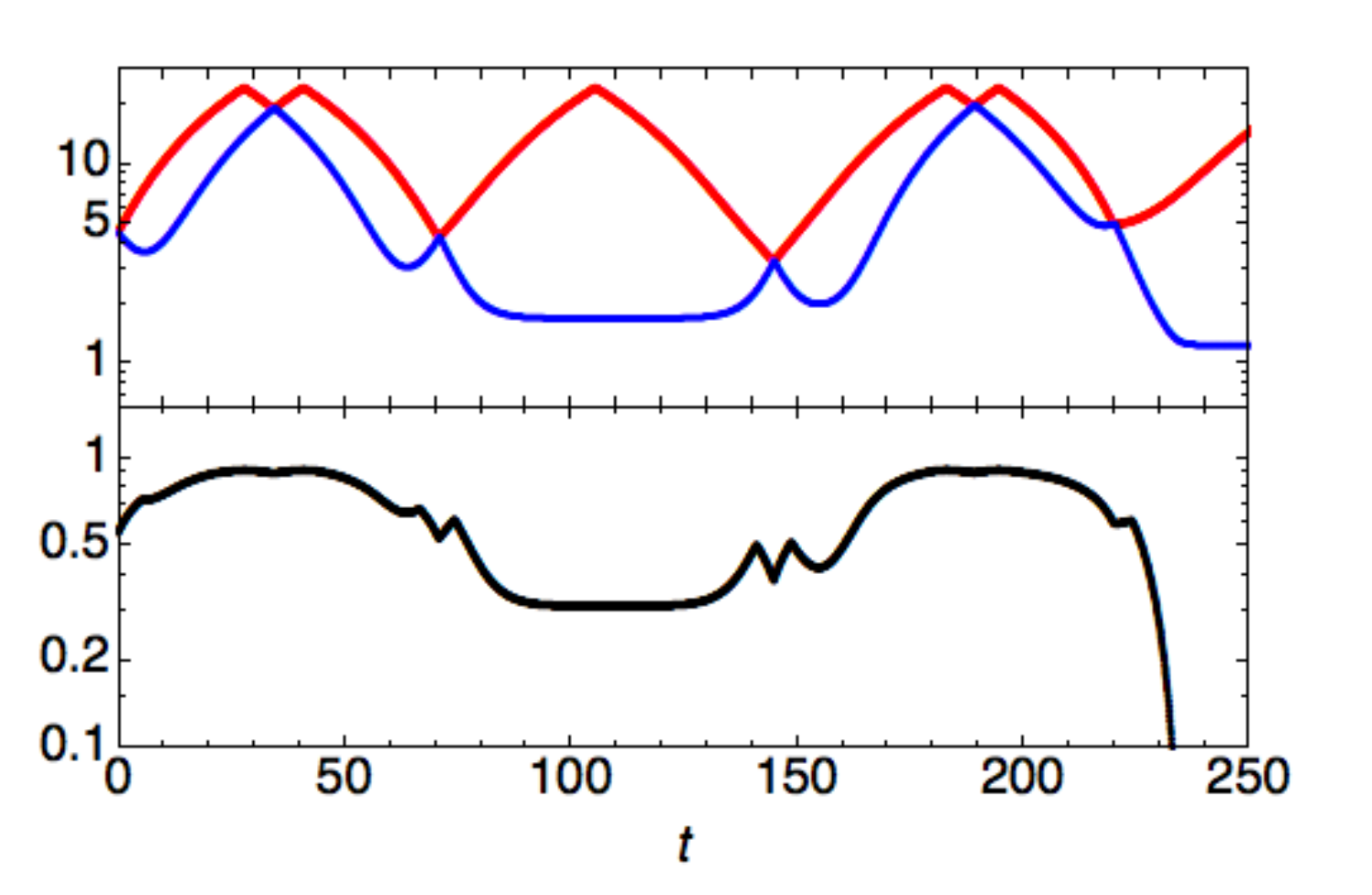}
\end{center}
\caption{Evolution of a double-shell spacetime, when confined to a box of radius $R_{\rm ext}=24$.
The upper panel shows shell positions as function of coordinate time $t$ as measured between shells, for $M_1=1, w=0.2, m=0.9, R_i=4.45825863$. Lower panel shows the {\it lowest} value of $1-2M_{1,2}/R_{1,2}$ computed at the surface of both shells.}
\label{fig:double}
\end{figure}
The potential $V$ felt by an isolated shell with positive $w$ is easily seen, from Eq.\eqref{potential}, to diverge negatively as $R\to 0,\,\infty$. Somewhere in between it will feature a peak. For certain choices of the parameters this maximum will be positive, thus restricting classical motion either to the interior region, which contains the origin, or to the exterior region. Now, considering two shells, each time they cross their effective potential suddenly changes. Therefore, if it is possible to choose parameters such that (i) the potential for the exterior shell has a (positive) peak exterior to the (also positive) peak of the potential for the interior shell, (ii) there is a finite region between the two peaks where both potentials are negative (allowing for classical motion) and (iii) the event horizons of the two non-flat regions are inside the turning points, then such a setup is likely to yield oscillatory motion.

One or more of the properties above may be destroyed after a number of shell-crossings. However,
by scanning the three-dimensional parameter space $\{w,m,M_2\}$, fixing the total ADM mass $M_1=1$ without loss of generality, we indeed found configurations satisfying the above conditions. An example of the shell radius as a function of coordinate time (measured by an observer located between shells) is shown in Fig.~\ref{fig:double_sustained}. This solution requires fine tuning of parameters; otherwise we find that, generically, upon each crossing the gravitational mass of the interior shell changes, leading to a progressive change in the crossing radius and eventually giving rise to collapse or to continued expansion. This suggests that such configurations are generically unstable~\cite{Javier}. 

\noindent{\bf {\em IV. Delayed collapse and critical phenomena.}}
When evolving in confined geometries, even single shells show a richer structure, without any need of fine-tuning:
the shell can collapse promptly, it can oscillate forever or it can bounce at the wall and then collapse.
This richer dynamics has a counterpart on the dynamics of two-shell systems, whose final outcome
depends sensitively on the initial conditions and parameters chosen.

\begin{figure}[h!]
\begin{center}
\includegraphics[width=0.5\textwidth]{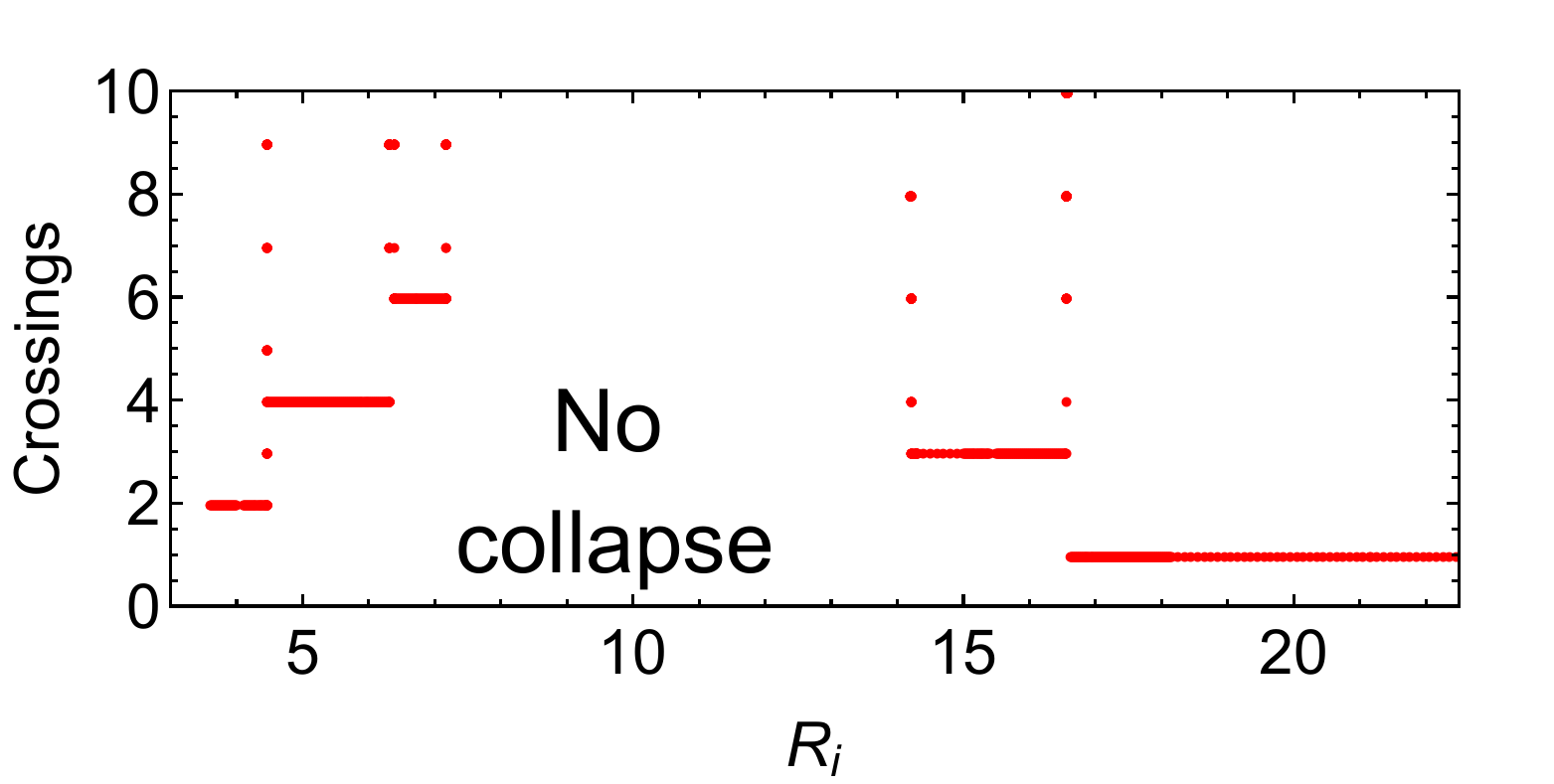}
\end{center}
\caption{Evolution of a double-shell spacetime with type A initial conditions, when confined to a box of radius $R_{\rm ext}=24$.
We find no collapse (or then it occurs only after a very large number of crossings not covered by our numerical results) for 
$7.2\lesssim R_i\lesssim 14.2$. We find arbitrarily large number of crossings at each critical point, where the number of crossings changes.}
\label{fig:crossings}
\end{figure}

An example concerning the evolution of initial data of type-A is shown in Fig.~\ref{fig:double}, for which the system only collapses after five crossings. Because the number of crossings is odd, it means that it is the initially outermost shell that eventually collapses. We find that the number of crossings, and even its parity, are highly sensitive to the initial conditions. This sensitiveness for type-A initial data is illustrated in Fig.~\ref{fig:crossings}, where we show the number of crossings between the two shells as a function of $R_i$.
There is a certain ``fractal-like'' structure not visible in the figure: the transition between one of the large plateaus to another (say, from $2\to 4$ crossings at $R_i\sim 4.4$) when zoomed-in looks like the Fig.~\ref{fig:crossings} itself, but starting with an odd number of crossings (3 in this case). The number of crossings increases, and the length of the corresponding plateau decreases as one approaches $R_i\sim 7.2$. Between $R_i\sim 7.2-14.2$, we find no collapse, but it settles in again for $R_i>14.2$.
Transitions that change parity are usually associated with a mass-gap (discontinuity) in the mass of the newly formed BH. 
For type-A initial data, we find that the odd-odd transitions (which are barely visible in the plot) are associated with ``critical'' points~\cite{critical_def}; this means that the mass enclosed by the horizon (when it first forms) is not a smooth function of the distance $R_i$. Some of the features we observe -- in particular decreasing plateaus and larger number of crossings between critical points -- are similar to the ones present in the collapse of scalars~\cite{Javier2} (see in particular Figs. 7-9 in Ref.~\cite{Buchel:2013uba}). This structure is highly dependent on the fine details of energy exchange between the shells:
changing the location of the boundary may change the outcome even at a qualitative level.

\begin{figure}[h!]
\begin{center}
\includegraphics[width=0.4\textwidth]{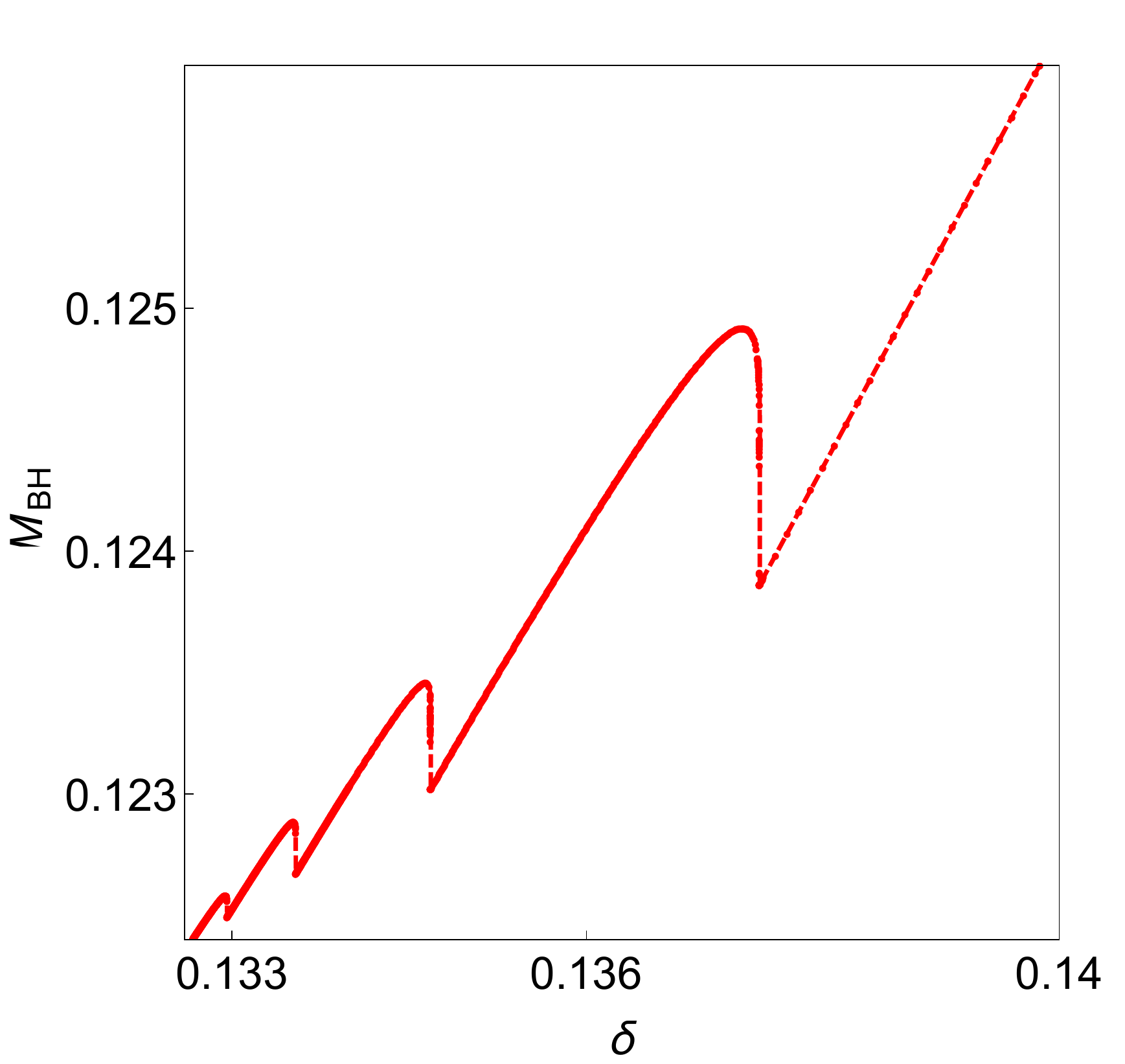}
\end{center}
\caption{BH mass as a function of the initial energy content $\delta$. From right to left, before each peak in the BH mass, the number of shell crossings is $1,3,5,7,9$. We do not observe any mass gap in the transitions. For $\delta\lesssim 0.13267$ we find no collapse.}
\label{fig:Mvsdelta_unequal}
\end{figure}
This type of behavior is more clearly seen for type-B initial data, controlled by the $\delta$ parameter.
The results are summarized in Fig.~\ref{fig:Mvsdelta_unequal}.

For $\delta \gtrsim 0.13267$ the system always collapses to a BH (For $\delta \geq 2$ an horizon encloses the initial configuration). Collapse is triggered by energy exchange between the shells, as indicated by the fact that collapse occurs only
after the shells cross once or multiple times. For these particular parameters, collapse occurs after an odd number of crossings, indicating that it is the initially outermost (and more energetic) shell that collapses. The dependence of the BH mass on $\delta$ is shown in Fig.~\ref{fig:Mvsdelta_unequal}. The BH mass $M_{BH}$ displays critical behavior. 
It is a continuous function of $\delta$, but its derivative is not $C^0$ at the critical points. In fact, the mass function is smooth to the right of the transition point, but its derivative blows up to the left of the transition point, signalling critical behavior~\cite{Gundlach:2002sx,Gundlach:2007gc}. Note that the final state is a BH on both sides of the critical point.
Each critical point is the limit of a branch of shell configurations which crossed a fixed number of times before
forming an horizon. Close to, and to the left of these critical points, the BH mass can be characterized by
\be
M_{BH}-M_0 \sim |\delta-\delta_*|^\gamma\,,\label{eq:critical_points}
\ee
For these particular initial configurations, we find no mass gap, and $M_0$ is the limiting BH mass at the critical point. The critical exponent is $\gamma\sim 0.2$ for the first transition (separating collapse after one and three shell crossings).

Our results are similar in many aspects to those describing gravitational collapse of scalar fields in AdS space or in spacetimes confined by artificial walls: nonlinear blueshift causes collapse and critical behavior is observed at each transition~\cite{Bizon:2011gg,Olivan:2015fmy,Okawa:2014nea} (notice the striking qualitative agreement between our Fig.~\ref{fig:Mvsdelta_unequal} and Fig. 1 in 
Ref.~\cite{Bizon:2011gg}, and most specially the overall consistency with Figs.7-9 in Ref.~\cite{Buchel:2013uba}).

\noindent{\bf{\em V. Conclusions.}}
Gravitation still hides many surprises. A two-shell, spherically symmetric system
is perhaps the simplest one can conceive of having interesting dynamics, described by {\it first order ODEs}. We have shown
that such setup displays a richness of phenomena, including oscillating solutions and critical phenomena,
but there is certainly a number of aspects that we have not yet explored. 

Although our setup is artificial, in that the confining wall was put by hand, we are 
confident that these features will be observed in asymptotically AdS spacetime,
where the timelike boundary naturally provides confinement~\footnote{After submission of this work, the generalization to asymptotically AdS spacetimes was studied in Ref.~\cite{Brito:2016xvw}.}. Single oscillating shells in these backgrounds exist, as long as a centrifugal barrier is present. When considering rotating shells this barrier is automatically generated~\cite{Delsate:2014iia}. In the non-rotating case, a positive pressure can have the same effect: this was studied recently~\cite{Mas:2015dra}, supporting our arguments that phenomena similar to what we described here will find its place in AdS\footnote{See also \cite{Rocha:2015tda} for oscillating rotating shells in higher dimensions without a cosmological constant.}. Thermalization on arbitrarily large timescales occurs in our setup; it would be interesting to frame these results within the gauge-gravity duality. Our study will hopefully help in shinning light over the mechanism(s) at play in the nonlinear instability and turbulent phenomena in AdS. Our particular setup is appealing because (i) our results show that similar phenomena occurs with (timelike) particles; (ii) it does so directly in position space,
whereas more traditional analysis focus on Fourier space (but see Ref.~\cite{Dimitrakopoulos:2014ada});
(iii) it does not require complex computational work and (iv) it can be generalized in a number of ways, to higher-dimensional frameworks, to asymptotically AdS, including shells with charge, rotation, etc.

\noindent{\bf{\em Acknowledgments.}}
%
We are indebted to Roberto Emparan, Carsten Gundlach, Javier Mas, David Mateos, Ken-ichi Nakao, Paolo Pani and Alexandre Serantes for their many useful comments and suggestions.
V.C. thanks the Departament de F\'{\i}sica Fonamental at Universitat de Barcelona, where this work was started, for hospitality.
V.C. acknowledges financial support provided under the European Union's H2020 ERC Consolidator Grant ``Matter and strong-field gravity: New frontiers in Einstein's theory'' grant agreement no. MaGRaTh--646597,
and FCT for Sabbatical Fellowship nr. SFRH/BSAB/105955/2014.
Research at Perimeter Institute is supported by the Government of Canada through Industry Canada and by the Province of Ontario through the Ministry of Economic Development $\&$
Innovation.
This work was supported by the H2020-MSCA-RISE-2015 Grant No. StronGrHEP-690904.
J.V.R. acknowledges financial support from the European Union's Horizon 2020 research and innovation programme under the Marie Sk\l{}odowska-Curie grant agreement No REGMat-656882.
J. V. R. was also partially supported by the Spanish MINECO under project FPA2013-46570-C2-2-P.

%

\bibliographystyle{h-physrev4}
\bibliography{Ref}

\begin{thebibliography}{10}

\bibitem{Cardoso:2014uka}
V.~Cardoso, L.~Gualtieri, C.~Herdeiro and U.~Sperhake,
\newblock Living Rev. Relativity {\bf 18}, 1 (2015), [1409.0014].

\bibitem{Dias:2015nua}
O.~J.~C. Dias, J.~E. Santos and B.~Way,
\newblock 1510.02804.

\bibitem{Lehner:2014asa}
L.~Lehner and F.~Pretorius,
\newblock Ann. Rev. Astron. Astrophys. {\bf 52}, 661 (2014), [1405.4840].

\bibitem{Sperhake:2008ga}
U.~Sperhake, V.~Cardoso, F.~Pretorius, E.~Berti and J.~A. Gonzalez,
\newblock Phys. Rev. Lett. {\bf 101}, 161101 (2008), [0806.1738].

\bibitem{Sperhake:2009jz}
U.~Sperhake {\em et~al.},
\newblock Phys. Rev. Lett. {\bf 103}, 131102 (2009), [0907.1252].

\bibitem{Sperhake:2012me}
U.~Sperhake, E.~Berti, V.~Cardoso and F.~Pretorius,
\newblock Phys. Rev. Lett. {\bf 111}, 041101 (2013), [1211.6114].

\bibitem{Healy:2015mla}
J.~Healy, I.~Ruchlin and C.~O. Lousto,
\newblock 1506.06153.

\bibitem{Choptuik:1992jv}
M.~W. Choptuik,
\newblock Phys. Rev. Lett. {\bf 70}, 9 (1993).

\bibitem{Gundlach:2002sx}
C.~Gundlach,
\newblock Phys. Rept. {\bf 376}, 339 (2003), [gr-qc/0210101].

\bibitem{1993gnsm.book.....C}
D.~{Christodoulou} and S.~{Klainerman},
\newblock {\em The global nonlinear stability of the Minkowski space}
  (Princeton University Press, Princeton, 1993).

\bibitem{Bizon:2011gg}
P.~Bizon and A.~Rostworowski,
\newblock Phys. Rev. Lett. {\bf 107}, 031102 (2011), [1104.3702].

\bibitem{Buchel:2012uh}
A.~Buchel, L.~Lehner and S.~L. Liebling,
\newblock Phys. Rev. {\bf D86}, 123011 (2012), [1210.0890].

\bibitem{Balasubramanian:2014cja}
V.~Balasubramanian, A.~Buchel, S.~R. Green, L.~Lehner and S.~L. Liebling,
\newblock Phys. Rev. Lett. {\bf 113}, 071601 (2014), [1403.6471].

\bibitem{Craps:2015jma}
B.~Craps and O.~Evnin,
\newblock {AdS (in)stability: an analytic approach},
\newblock in {\em {21st European String Workshop: The String Theory Universe
  Leuven, Belgium, September 7-11, 2015}}, 2015, [1510.07836].

\bibitem{Abajo-Arrastia:2014fma}
J.~Abajo-Arrastia, E.~da~Silva, E.~Lopez, J.~Mas and A.~Serantes,
\newblock JHEP {\bf 05}, 126 (2014), [1403.2632].

\bibitem{Okawa:2014nea}
H.~Okawa, V.~Cardoso and P.~Pani,
\newblock Phys. Rev. {\bf D90}, 104032 (2014), [1409.0533].

\bibitem{Okawa:2015xma}
H.~Okawa, J.~C. Lopes and V.~Cardoso,
\newblock 1504.05203.

\bibitem{Olivan:2015fmy}
D.~S. Oliv\'an and C.~F. Sopuerta,
\newblock 1511.04344.

\bibitem{Israel:1966rt}
W.~Israel,
\newblock Nuovo Cim. {\bf B44S10}, 1 (1966),
\newblock [Nuovo Cim.B44,1(1966)].

\bibitem{Dray:1985yt}
T.~Dray and G.~'t~Hooft,
\newblock Commun. Math. Phys. {\bf 99}, 613 (1985).

\bibitem{Nunez:1993kb}
D.~Nunez, H.~P. de~Oliveira and J.~Salim,
\newblock Class. Quant. Grav. {\bf 10}, 1117 (1993), [gr-qc/9302003].

\bibitem{Lake:1979zz}
K.~Lake,
\newblock Phys. Rev. {\bf D19}, 2847 (1979).

\bibitem{Darmois:1927}
G.~Darmois,
\newblock M\'emorial de Sciences Math\'ematiques {\bf fascicule 25}, 1 (1927).

\bibitem{Hawking:1973uf}
S.~W. Hawking and G.~F.~R. Ellis,
\newblock {\em {The Large Scale Structure of Space-Time}}Cambridge Monographs
  on Mathematical Physics (Cambridge University Press, 2011).

\bibitem{1999PThPh_v2}
K.~{Nakao}, D.~Ida and N.~Sugiura,
\newblock Progress of Theoretical Physics {\bf 101}, 47 (1999).

\bibitem{1999PThPh.101..989I}
D.~{Ida} and K.~{Nakao},
\newblock Progress of Theoretical Physics {\bf 101}, 989 (1999).

\bibitem{Javier}
{This conclusion concerns only free shells. As we will show, once a mirror is
  added to the system, eternally oscillating configurations exist for generic
  conditions, and we therefore suspect that stable solutions also exist. This
  means, in particular, that stable solutions describing oscillating pairs of
  shells exist in asymptotically AdS spacetimes.}

\bibitem{critical_def}
{We define critical points to be points at which the number of crossings
  changes. These are also critical in the usual way, in that the BH mass
  function is not $C^1$ but is described by \eqref{eq:critical_points}.}

\bibitem{Javier2}
{We thank Javier Mas for bringing this point to our attention, and for sharing
  unpublished results which support further this behavior between transitions.}

\bibitem{Buchel:2013uba}
A.~Buchel, S.~L. Liebling and L.~Lehner,
\newblock Phys. Rev. {\bf D87}, 123006 (2013), [1304.4166].

\bibitem{Gundlach:2007gc}
C.~Gundlach and J.~M. Martin-Garcia,
\newblock Living Rev. Rel. {\bf 10}, 5 (2007), [0711.4620].

\bibitem{Brito:2016xvw}
R.~Brito, V.~Cardoso and J.~V. Rocha,
\newblock 1602.03535.

\bibitem{Delsate:2014iia}
T.~Delsate, J.~V. Rocha and R.~Santarelli,
\newblock Phys. Rev. {\bf D89}, 121501 (2014), [1405.1433].

\bibitem{Mas:2015dra}
J.~Mas and A.~Serantes,
\newblock Int. J. Mod. Phys. {\bf D24}, 1542003 (2015), [1507.01533].

\bibitem{Rocha:2015tda}
J.~V. Rocha,
\newblock Int. J. Mod. Phys. {\bf D24}, 1542002 (2015), [1501.06724].

\bibitem{Dimitrakopoulos:2014ada}
F.~V. Dimitrakopoulos, B.~Freivogel, M.~Lippert and I.-S. Yang,
\newblock JHEP {\bf 08}, 077 (2015), [1410.1880].

\end{thebibliography}

\end{document}